\documentclass{iau} 
\usepackage{graphicx}
\usepackage{bm}
\usepackage[usenames, dvipsnames]{color}                                                                  
\usepackage{amsmath}

\usepackage{url}
\usepackage{hyperref}
\hypersetup{
    final=true,
    pageanchor=true,
    colorlinks=true,
    breaklinks=true,
    linkcolor=blue,
    citecolor=blue,
    urlcolor=blue,
    pdfpagemode=UseNone,
    pdftitle={sTools - a data reduction pipeline for GFPI and HiFI},
    pdfauthor={Kuckein et al.},
    pdfsubject={Solar Physics},
    pdfkeywords={techniques: image processing, methods: data analysis, astronomical data bases: miscellaneous}}

\newcommand*\farcs{\ensuremath{\overset{\prime\prime}{.}}}

\title[sTools -- a data reduction pipeline for GFPI and HiFI] %% give here short title %%
{sTools -- a data reduction pipeline for the GREGOR Fabry-P{\'e}rot Interferometer \\
and the High-resolution Fast Imager \\
at the GREGOR solar telescope}

\author[Kuckein \etal]   %% give here short author list %%
       {C. Kuckein$^1$,
        C. Denker$^1$,
        M. Verma$^1$,
        H. Balthasar$^1$, \\
        S. J. Gonz{\'a}lez Manrique$^{1,2}$,
        R. E. Louis$^1$, \and
        A. Diercke$^{1,2}$}

\affiliation{$^1$Leibniz-Institut f{\"u}r Astrophysik Potsdam, An der Sternwarte 16, 14482, Potsdam, Germany \\ 
emails: {\tt ckuckein@aip.de, cdenker@aip.de} \\
$^2$ Universit{\"a}t Potsdam, Institut f{\"u}r Physik and Astronomie, 14476 Potsdam, Germany\\}
\pubyear{2017}
\volume{327}  %% insert here IAU Symposium No.
\jname{Fine Structure and Dynamics of the Solar Atmosphere}
\editors{S. Vargas Dom{\'i}nguez, A. G. Kosovichev, L. Harra, P. Antolin, eds.}
\begin{document}

\maketitle

\begin{abstract}
A huge amount of data has been acquired with the GREGOR Fabry-P{\'e}rot Interferometer (GFPI), 
large-format facility cameras, and since 2016 with the High-resolution Fast Imager (HiFI). 
These data are processed in standardized procedures with the aim of providing science-ready 
data for the solar physics community. For this purpose, we have developed a user-friendly data 
reduction pipeline called ``sTools'' based on the Interactive Data Language (IDL) and licensed 
under creative commons license. 
The pipeline delivers reduced and image-reconstructed data with a minimum of user interaction. 
Furthermore, quick-look data are generated as well as a webpage with an overview of the 
observations and their statistics. All the processed data are stored online at the GREGOR GFPI and 
HiFI data archive of the Leibniz Institute for Astrophysics Potsdam (AIP). 
The principles of the pipeline are presented together with selected high-resolution spectral 
scans and images processed with sTools.
\end{abstract}

\firstsection % if your document starts with a section,
              
% ------------------------------------------------------------------------------------------   
\section{Introduction}
% ------------------------------------------------------------------------------------------

The largest and most powerful solar telescopes are ground-based facilities. However, compared to their counterparts in space such as the Solar 
Dynamics Observatory (\cite[SDO; Pesnell et al. 2012]{sdo}), they often lack in providing user-friendly data pipelines to reduce the raw data.
Only few ground-based instruments have such a pipeline (\cite[e.g., de la Cruz Rodr{\'{\i}}guez \etal\ 2015]{crispred}). 
The data reduction process is often much more complex than for space missions and differs substantially 
from instrument to instrument. The need for 
automatized data reduction pipelines to produce science-ready data becomes crucial in the era of big data. 
Therefore, in the framework of the EU-funded 
SOLARNET project, we developed a pipeline called ``sTools'' to reduce and prepare data acquired with the GREGOR Fabry-P{\'e}rot Interferometer 
(\cite[GFPI; Puschmann \etal\ 2012]{puschmann2012}), large-format imaging cameras, and the High-resolution Fast Imager (HiFI) located at 
Europe's largest solar telescope GREGOR (\cite[Schmidt \etal\ 2012]{gregor}). In Sect. 2 we will briefly describe the instruments supported
by the pipeline. Section 3 is dedicated to the description of the pipeline, and Sect. 4 reports on the data archive and provides an outlook.

% ------------------------------------------------------------------------------------------
\section{Instruments}
% ------------------------------------------------------------------------------------------

The GFPI is the successor of the G\"ottingen Fabry-P\'erot Interferometer (\cite[Bendlin \etal\ 1992]{bendlin1992},
\cite[Puschmann \etal\ 2006]{puschmann2006}) which was attached to the Vacuum Tower Telescope (VTT) at the Observatorio del Teide and is
now installed at GREGOR.
It is an imaging spectropolarimeter which operates in the visible and near infrared spectral range. The GFPI has two tunable etalons 
in a collimated setup. 
%Hence, the data reduction needs to account for the blueshift across the field-of-view (FOV). 
The two CCD
cameras, one for the broad-band and another for the narrow-band images, have the same image scale and a size of $1376\times1040$ pixels. 
The spatial scale of GFPI was derived comparing GFPI broad-band images from the year 2014 with continuum images from SDO which results
in 0\farcs0405\,pixel$^{-1}$. Hence, the field-of-view (FOV) of the GFPI in spectroscopic mode is $55\farcs7 \times 41\farcs5$. 

The blue imaging channel (BIC) of the GFPI is fed by light in the visible below $\sim$5300\,\AA. A beamsplitter allows for simultaneous image 
acquisition at two different wavelengths with two cameras. 
The observer aims for high-cadence with these cameras to minimize atmospheric seeing effects and assure high-quality
image reconstruction. Initially, two pco.4000 CCD cameras with a size of $4008 \times 2672$ pixels and a spatial sampling of 
0\farcs0347\,pixel$^{-1}$ were used. In 2016, these cameras were replaced by two synchronized sCMOS imagers
and the instrument is now named High-resolution Fast Imager (HiFI). Both cameras 
write their images into the same file. This requires the same exposure times for both cameras which can be achieved 
using neutral density filters. 
The HiFI chips have a size of $2560 \times 2160$ pixels and a spatial sampling of 0\farcs0253\,pixel$^{-1}$.
Of special interest is the frame rate of these cameras which is 49\,Hz or 98\,Hz with the full or a 
$1920 \times 1080$ pixels FOV, respectively. HiFi can be used independently from the GFPI, e.g., as the context image 
of the GREGOR Infrared Spectrograph (\cite[GRIS, Collados \etal\ 2012]{gris}).

% ------------------------------------------------------------------------------------------
\section{sTools pipeline}
% ------------------------------------------------------------------------------------------

A careful acquisition of calibration data each day is crucial to produce high-quality science data after the reduction process. 
The following calibration data are required: dark, flat field, resolution target, and
several pinhole images. In addition, the spectroscopic mode of the GFPI requires a long scan to derive the prefilter curve
of the interference filter. It is crucial to have dark images with the same exposure times as all the acquired data. 

All routines are written in the Interactive Data Language (IDL)
and are documented in the header. Moreover, the routines are strictly named with the prefix ``stools'' in order to avoid overlapping with 
other IDL routines. GFPI and HiFI data are recorded in the native format of DaVis, a software package from the company LaVision GmbH, G\"ottingen.

% ------------------------------------------------------------------------------------------
\subsection{GREGOR Fabry-P{\'e}rot Interferometer}
% ------------------------------------------------------------------------------------------

We will focus on the data reduction of the imaging spectroscopic mode. All computed images and parameters for the 
data calibration are stored in one single IDL save file. Single variables are added to or are extracted from IDL 
save files with special sTools routines.
To initialize the pipeline, the user only needs to specify the in and output directories,
the observed wavelength, and the on-chip binning of the cameras. 
First, the average dark images are computed for both cameras. 
In the next step the flat-field images are averaged for each position of the etalon. 
Other valuable information such as the positions of the etalon, accumulations, and step size is also saved. 
The pinhole and resolution target images are corrected for dark and flat-field. 
They are then used to derive the alignment parameters between both GFPI cameras such as rotation, displacement, pivot point, and 
magnification. The output is stored in an IDL structure. 

The blueshift correction needs to be performed across the FOV of the narrow-band images. 
For this task, flat-field images are taken and interpolated to a narrower and equidistant wavelength grid. 
The resulting spectral profiles for each pixel are slightly smoothed by convolving them with a normalized kernel of $(1,2,4,8,4,2,1)$ units. 
A polynomial fit of second order to the line core of the spectral line yields the central position on the wavelength axis. 
We repeat this step for all spectra across the FOV. The average of all central positions is then computed and subtracted
from each individual position. Hence, we obtain the displacement in wavelength, i.e., the blueshift correction, for all pixels within the FOV.
The correction is applied by displacing the spectra using cubic spline interpolation. 
For the narrow-band camera, the aforementioned flat field can now be used together with the blueshift correction to compute a 
normalized master flat, i.e., a single two-dimensional gain table, where the spectral information is removed.
The user can choose either this master flat or the former spectral-dependent flat field to correct the data.

\begin{figure}[!t]
 %   \begin{center}
    \begin{minipage}{.5\textwidth}
 %       \begin{center}
         \includegraphics[width=0.99\textwidth]{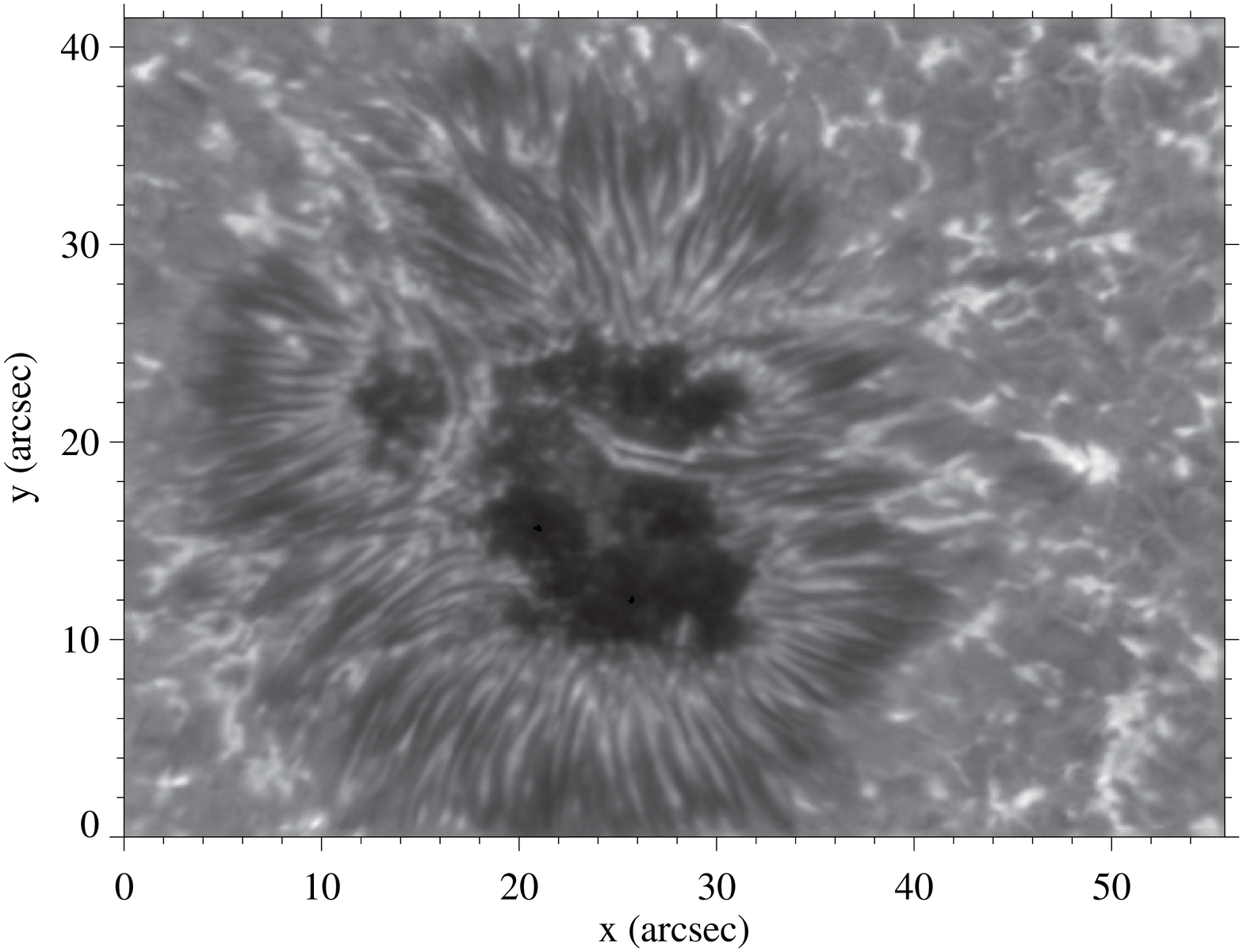}
        %        \end{center}
    \end{minipage}%
    \begin{minipage}{0.5\textwidth}
%        \begin{center}
        \includegraphics[width=0.99\textwidth]{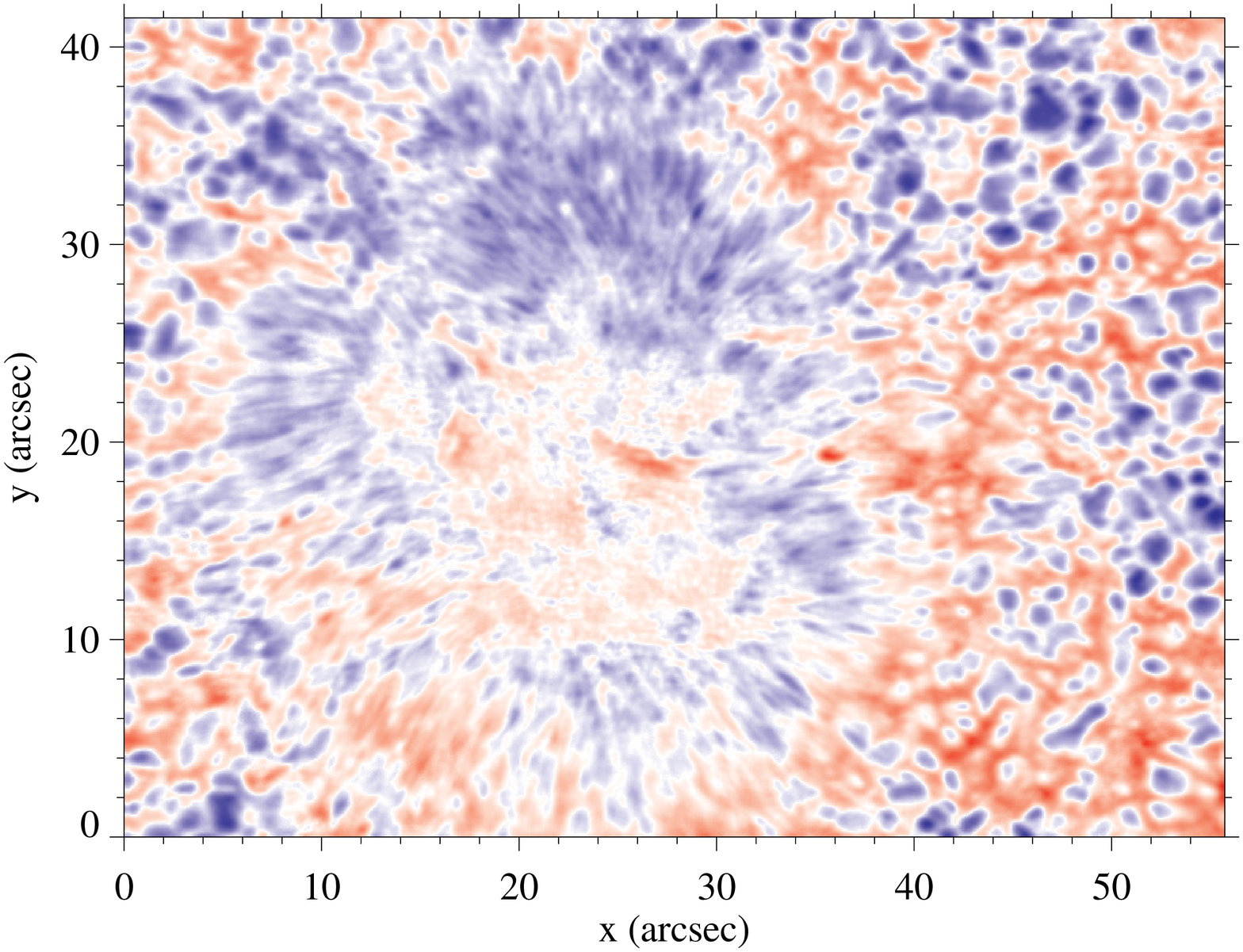}
%        \end{center}
    \end{minipage}   
 %   \end{center}
    \caption{\textit{Left}: MOMFBD-restored GFPI image of active region NOAA 12121 in the line core of Fe\,\textsc{i}~6173\,\AA\ taken on 
    2014 July 27. \textit{Right}: LOS velocities clipped at $\pm 1$\,km\,s$^{-1}$.}
   \label{fig1}
\end{figure}

The last step before correcting the science observations is to derive the prefilter curve of the GFPI. 
To this end, we use a long scan with a small step size covering a large fraction of the etalon's scan range. 
This scan is ideally taken at the solar disk center under quiet-Sun conditions to avoid spectral line shifts. 
After dark, flat-field, and blueshift corrections a GFPI mean spectrum, which represents the average transmission profile
of the narrow-band interference filter, is computed. To extract the trend of the prefilter curve we 
take advantage of the Fourier Transform Spectrometer (\cite[FTS; Neckel \& Labs 1984]{fts}) spectrum from the Kitt Peak National Observatory.
The FTS spectrum is first convolved with the theoretical GFPI transmission profile and then matched to the observed GFPI mean spectrum by minimizing
the wavelength offset. It follows a resampling of the FTS spectrum to the GFPI wavelength scale. Afterwards, the ratio of both 
spectra is computed. Finally, the ratio, with the exception of parts showing large variations, is fitted using a double Gaussian. 
Linear least-squares fitting between model and observations is performed with
the MPFIT package (\cite[Markwardt 2009]{markwardt2009}). 

All calibration data are now ready to be applied to the science data. The pipeline splits now into two different branches. On 
the one hand, quick-look physical maps are being generated, e.g., line-core intensity images, 
line-of-sight (LOS) velocity maps, equivalent width maps, seeing quality parameters, etc. The results are stored into an IDL structure
together with the reduced broad- and narrow band images (Level 1.0 data). 
On the other hand, the data are prepared for image-restoration (Level 2.0 data) with 
Multi-Object Multi-Frame Blind Deconvolution (\cite[MOMFBD; L{\"o}fdahl 2002]{momfbd}, 
\cite[van Noort \etal\ 2005]{vannoort2005}).
The restored images are finally written in the Flexible Image Transport System format (\cite[FITS, Wells \etal\ 1981]{fits}). 
One example of a restored line-core image and its associated LOS velocity map is shown in Fig. \ref{fig1}.

% ------------------------------------------------------------------------------------------
\subsection{High-resolution Fast Imager}
% ------------------------------------------------------------------------------------------

\begin{figure}[!t]
 %   \begin{center}
    \begin{minipage}{.5\textwidth}
 %       \begin{center}
        \includegraphics[width=0.99\textwidth]{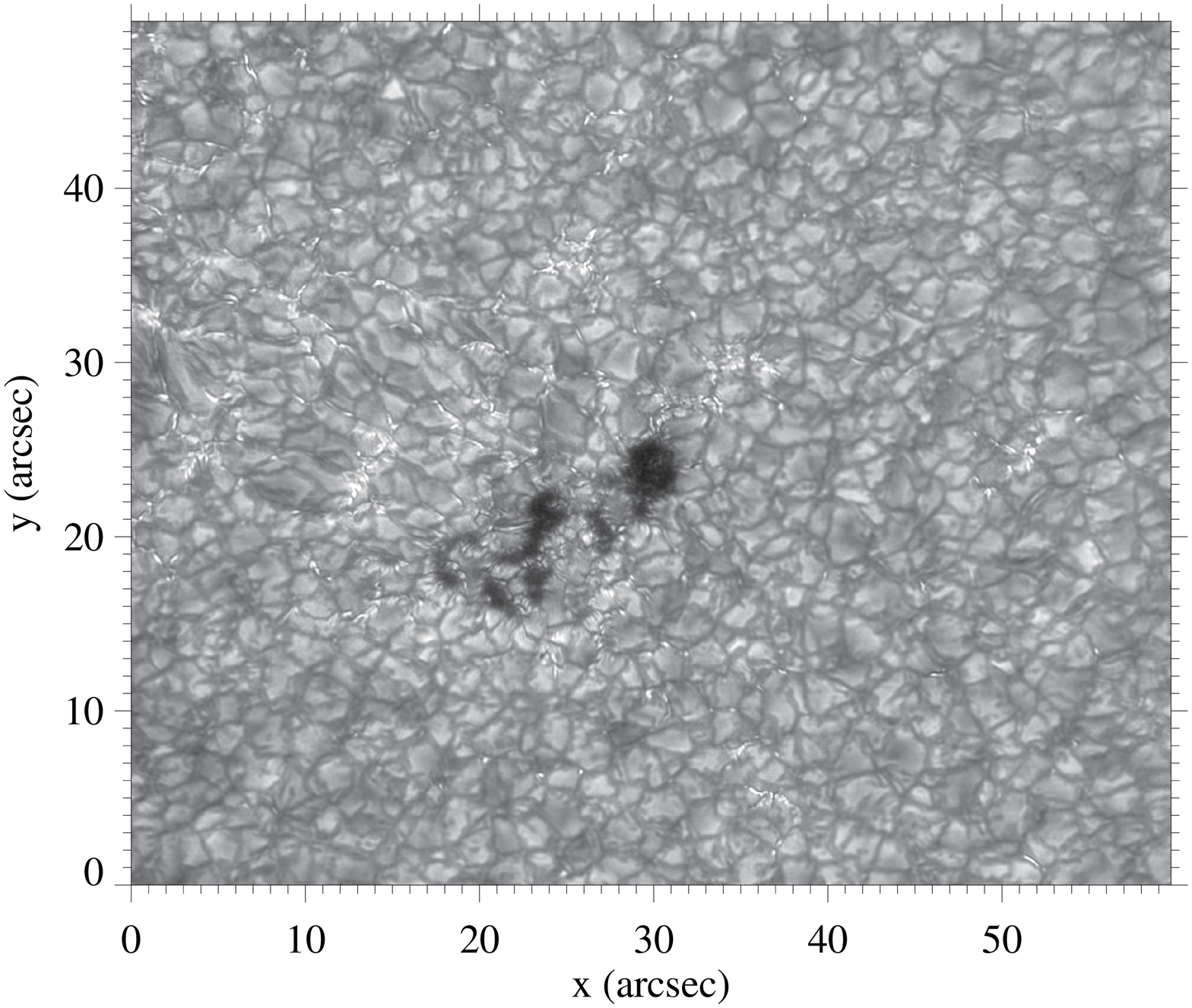}
%        \end{center}
    \end{minipage}%
    \begin{minipage}{0.5\textwidth}
%        \begin{center}
        \includegraphics[width=0.99\textwidth]{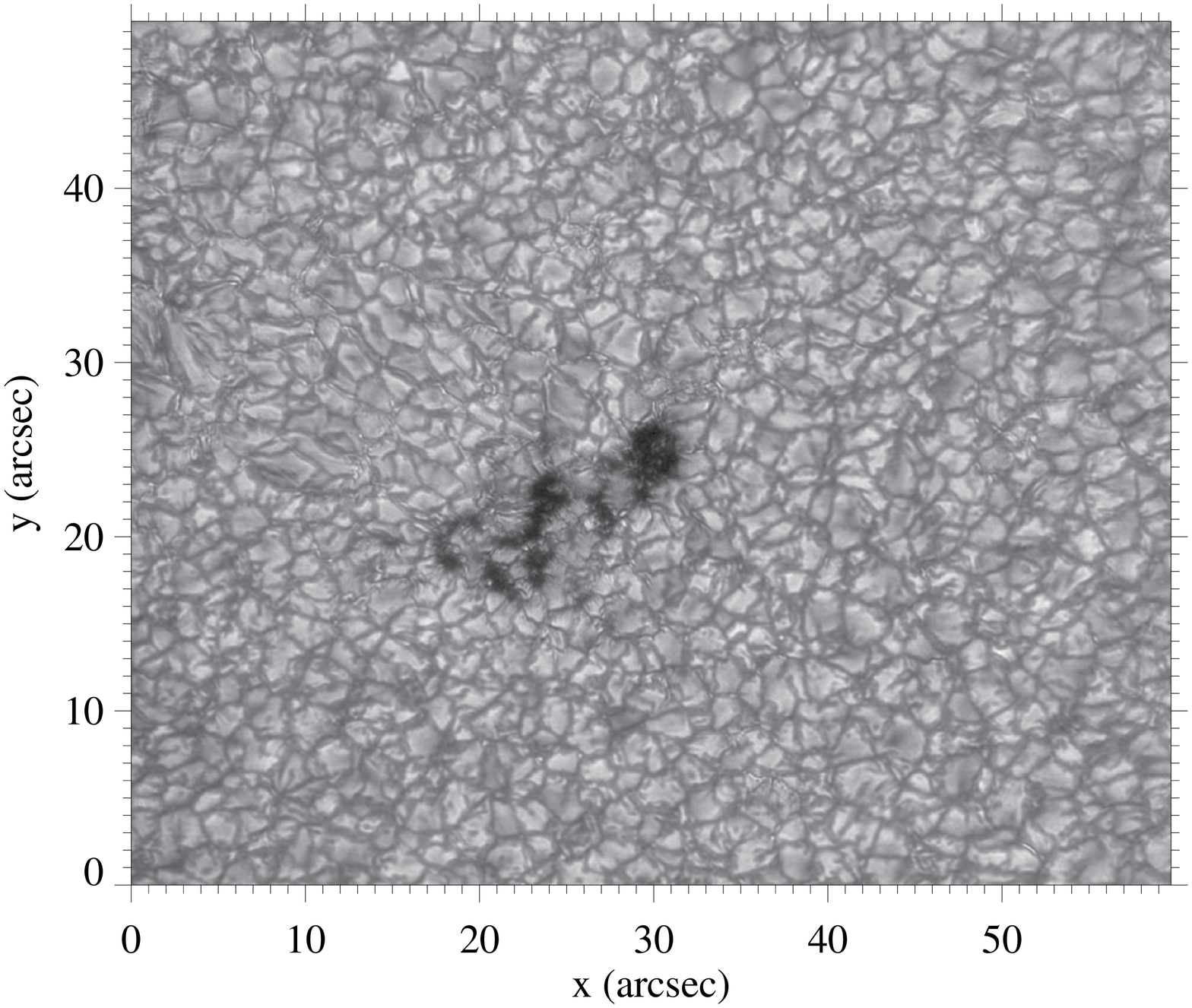}       
%        \end{center}
    \end{minipage}   
 %   \end{center}
    \caption{\textit{Left} to \textit{right}: G-band (4307\,\AA) and blue-continuum (4506\,\AA) speckle-reconstructed HiFI images of 
      active region NOAA 12530 observed on 2016 April 11 at GREGOR.}
   \label{fig2}
\end{figure}

We follow a similar approach for the HiFi data reduction. The user needs to introduce the directory of the observations, 
telescope and user names, and the two wavelengths of the respective interference filters. The average dark, 
flat field, pinholes, and resolution target images are then computed and saved as individual FITS files. The file name has a two-digit
suffix which identifies the type of calibration data. Moreover, the FITS header includes basic information 
such as date and time of the observations, wavelength, telescope, dimensions of the images, etc.  

The target frames are used to align both cameras. The obtained 
parameters such as displacement, rotation, and magnification are stored in an IDL save file. Due to the setup of the cameras, 
both have the same image scale, however, there might be a slight displacement in the vertical and horizontal axes.
The science observations are then corrected for dark and for artifacts on the chip by using the flat-field images. If more than one
burst of flat-field images was taken, the pipeline chooses the one closest in time to the science observations. 

An image quality check to sort the images is performed at this point. 
The reason for this is that HiFI, with its high frame rate, acquires more images than actually needed for the image restoration. 
Usually 500 images are taken per burst and only the 100 best are used for the restoration. The other images are dropped. 
However, the user can change this criterion. 
We take advantage of the Median Filter-Gradient Similarity (MFGS) introduced by \cite[Deng \etal\ (2015)]{deng2015}, to 
scrutinize the image quality. 
The MFGS code is implemented in an independent IDL routine and is applied to every single image. Each pixel has now a value between 0 and 1. 
This value and additional statistics related to the MFGS are stored in a separate IDL save file and also added to the FITS header.
Finally, the average MFGS value over the whole image is used as a reference. 
Note that the MFGS value is only calculated for the images of one of the cameras. This is justified 
since both cameras strictly acquire images at the same time, hence, both were recorded under the same seeing conditions. 

Finally, the sorted images according to their MFGS value (Level 1.0 data) are written into a FITS file using standard routines of the 
IDL Astronomy User's Library hosted at \mbox{\href{http://idlastro.gsfc.nasa.gov}{idlastro.gsfc.nasa.gov}}.
The two bursts of images are written alternately in a single FITS file using image extensions (\cite[Ponz \etal\ 1994]{ponz1994}). 
Odd extension numbers belong to images from one filter
whereas even numbers belong to the other filter. The file has a detailed primary header with all relevant information about the
observations. Furthermore, there are extension headers for each image showing only image specific information, e.g., the wavelength, 
image statistics, and MFGS information. 
The data restoration is carried out using the speckle-interferometry code KISIP (\cite[W{\"o}ger \& von der L{\"u}he 2008]{kisip})
to produce science-ready data (Level 2.0 data) as shown in Fig. \ref{fig2}.

% ------------------------------------------------------------------------------------------
\section{Data archive and outlook}
% ------------------------------------------------------------------------------------------

GFPI and HiFI data are stored at the data archive of AIP (\href{http://gregor.aip.de}{gregor.aip.de}). In addition, the pipeline generates a webpage, which includes an overview of the acquired observations as well
as quick-look images and statistics. 
The data recorded since 2014 will be made public to the solar community. The sTools pipeline is under constant development. 
An official version will be released in 2017 via a version control system on the abovementioned webpage under creative commons license.

% ------------------------------------------------------------------------------------------
\begin{acknowledgements}
\noindent This work was carried out as a part of the SOLARNET project supported by the European 
Commission's 7th Framework Programme under grant agreement No. 312495.
The 1.5-meter GREGOR solar telescope was built by a German consortium under the leadership of the 
Kiepenheuer-Institut f\"ur Sonnenphysik in Freiburg (KIS) with the Leibniz-Institut f\"ur Astrophysik Potsdam (AIP), 
the Institut f\"ur Astrophysik G\"ottingen (IAG), the Max-Planck-Institut f\"ur Sonnensystemforschung in 
G\"ottingen (MPS), and the Instituto de Astrof\'isica de Canarias (IAC), and with contributions by the Astronomical 
Institute of the Academy of Sciences of the Czech Republic (ASCR). SJGM is grateful for financial support from the Leibniz Graduate School
for Quantitative Spectroscopy in Astrophysics, a joint project of AIP
and the Institute of Physics and Astronomy of the University of Potsdam.
\end{acknowledgements}

% ------------------------------------------------------------------------------------------

\end{document}